# A Review On Securing Distributed Systems Using Symmetric Key Cryptography

Ramesh Babu[1], George Abraham[2] and Kiransinh Borasia[3]

[1,2,3]School of Computing Science and Engineering, VIT University, Vellore, India
[1]krameshbabu@vit.ac.in, [2]george.abraham2011@vit.ac.in, [3]borasiakiransinh.ranjitsinh2011@vit.ac.in

## Abstract

*This review is aimed to evaluate the importance of Symmetric Key Cryptography for Security in Distributed Systems. Businesses around the world as well as research and other such areas rely heavily on distributed systems these days. Hence, security is also a major concern due to the openness of the system. Out of the various available security measures, we, in this paper, concentrate in general on the symmetric key cryptographic technique. We review two widely used and popular symmetric key cryptographic algorithms, viz. DES and AES. These two algorithms are evaluated on the parameters such as key size, block size, number of iterations, etc.*

## 1. Introduction

Distributed Systems have grown rapidly over the last two decades. A distributed system consists of multiple autonomous computers that communicate with each other through a computer network in order to achieve a common goal. This system is currently being used for several commercial, educational and industrial applications, where transactions are taking place through an online client-server setup, fileservers and tele-networks. The data or the systems are not within the control of one person; it is spread over a network. This data transfer between client and server requires a high priority security measure as the data is left open in the communication [9][13]. In such a widely used environment, security is the main concern. Some robust security measures should be undertaken to ensure that the systems and data are well secured.

Security in distributed systems focuses on both the hardware and software areas. The hardware security focusses on the keys, locks, cards, monitoring the visitations, etc. The main research work is done in software security. The software security deals with:

- *Authentication* [4][2]: Verifying the identity of the user
- *Confidentiality* [3][8][6]: Message transmitted must be read only by the intended users and not by anyone else.
- *Authorization* [4][2]: Verifying if the user has the adequate permission to use the service
- *Integrity* [3][8][6]: Message that is transmitted over the network must be similar to the one that is sent and not tampered with.
- *Availability* [12][3]: Even in case of any failures, the system must be available
- *Accountability* [10][3]: All user actions that is security critical must be traceable back to the user

Some of the common approaches for ensuring security in distributed systems are Cryptography, Digital Signatures, SSL, etc. Digital Cryptography provides the basic computer security mechanism. It is a method by which the message is encoded in such a format that only the intended recipient can decode and understand it. This technique is implemented mainly to provide for authentication, confidentiality and integrity in distributed systems [1]. An encrypted data cannot be ideally tampered with, thereby preserving the integrity of the message.

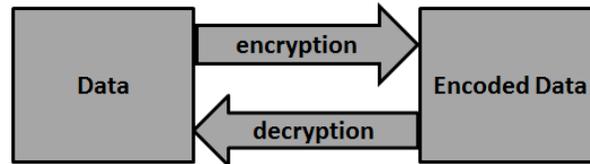

Figure 1. Basic Cryptography

Cryptography, based on the way it is implemented is divided into two categories: symmetric key cryptography and asymmetric key cryptography. Symmetric Key Cryptography is one in which both sender and receiver uses the same shared secret key for both encryption and decryption. In Asymmetric Key Cryptography, the sender and receiver use two keys – private key and public key for encryption and decryption. This also leads to the overhead of ensuring a proper key management system. In this paper, we discuss about providing security to distributed systems with the help of symmetric key cryptographic algorithms.

## 2. Symmetric Key Cryptography

Symmetric Key Cryptography [1][7][6][11] is a class of cryptography techniques where identical cryptographic keys are used for encoding of plain text as well as decoding of cipher text. The common key is a shared secret between two or more parties. Symmetric key encryption is also called as secret key, single key, shared key, one-key and private key encryption. A substitution cipher replaces one symbol with another. If the symbols in the plaintext are alphabetic characters, we replace one character with another. The simplest substitution cipher is a shift cipher (additive cipher).

Symmetric key encryption can use either stream ciphers or block ciphers [11]. Stream ciphers take the message one bit at a time and encrypt it. Block ciphers take a block of bits of the message and encrypt it. Symmetric Key Cryptography is generally used for long messages.

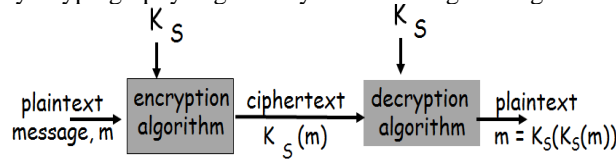

Figure 2. Symmetric Key Cryptography

There are many types of symmetric key cryptography algorithms – DES, AES, RC4, 3DES, TwoFish, etc. Each of the algorithms mainly differs with respect to the block size, the key size, the number of rounds the algorithm runs, etc.

### 8.1. Data Encryption Standard

The Data Encryption Standard (DES) algorithm was once a popular algorithm, when it was launched in the early 70's till late 90's when the National Institute of Standards and Technology has withdrawn DES as a security standard [5]. DES is a block cipher based on the Feistel structure and has a block size of 64 bits and an effective key of 56 bits to aid in transformation [5]. The overall view of the algorithm, the iteration block structure and its general scheme are shown below in Figure 3, Figure 4 and Figure 5, respectively.

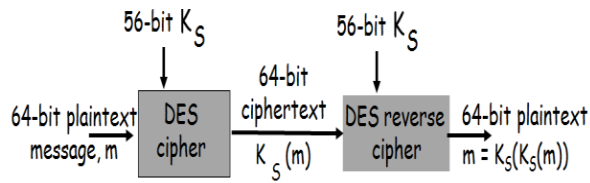

Figure 3. DES Algorithm – Overview

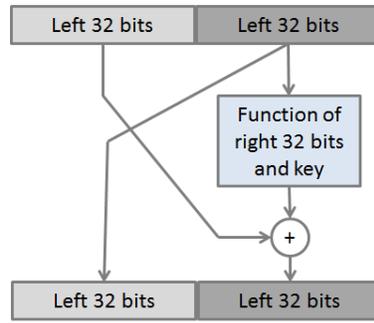

Figure 4. Iteration Block

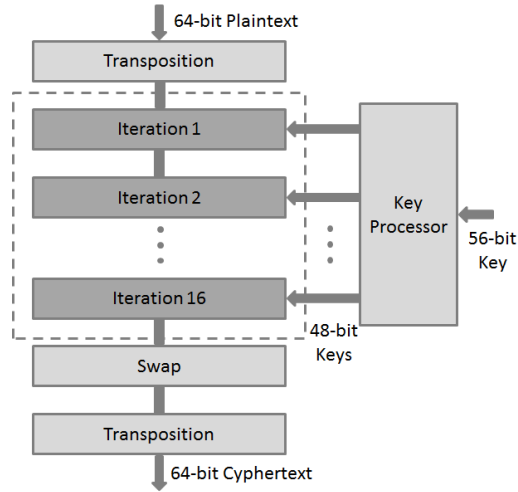

Figure 5. General DES Scheme

## 8.1. Advanced Encryption Standard

The Advanced Encryption Standard (AES) is a key-iterated block cipher [4] that has been approved by the U.S. Government [1]. The AES is originally the Rijndael Algorithm, created by Joan Daemen and Vincent Rijmen, as a part of the US National Institute of Standards and Technology (NIST) requirement to have an Advanced Encryption Standard to replace the obsolete Data Encryptions Standards (DES) algorithms. AES does not use a Feistel network and is based on a design principle known as a substitution-permutation network. It is fast in both software and hardware. The overview of the AES algorithm and its general scheme is shown in Figure 6 and Figure 7, respectively.

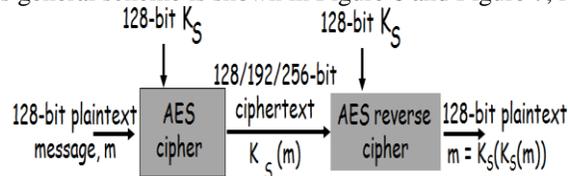

Figure 6. AES Algorithm – Overview

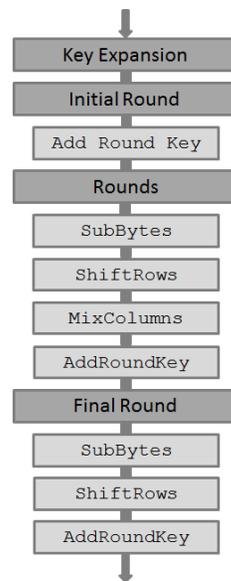

Figure 7. General AES Scheme

## 3. Existing Work

S. Distefano and A. Pulliafito [6], in their paper, talks about providing security for distributed systems by combining the asymmetric cryptography technique (RSA) with the symmetric cryptography technique (AES). The authors justify themselves for combining both the techniques due to the reason that using symmetric algorithms with key-splitting technique, alone has reliability issues like the key server being unavailable. They use AES for encrypting the message, as symmetric key approach is the best technique for data security. Then they encrypt the symmetric key using RSA algorithm to ensure security of the encryption key. They have implemented the algorithm into a service of the gLite Grid middleware, which they have named as the Grid Secure Storage System (GS3), as a file system ensuring the protection of both the data and the files.

S. V. Wunnava and E. Lule [12], in their paper, presents the design of a hardware based security processor module for network-connected systems. The security processor is implemented on the Data Encryption Standard (DES) algorithm using Very high speed integrated circuit Hardware Description Language (VHDL). They go for the creation of a hardware module because, as per their research, cryptographic algorithms implemented on Field Programmable Gate Arrays (FPGAs) and Complex Programmable Logic Devices (CPLDs) provide a great degree of flexibility, when compared to Application Specific Integrated Circuits (ASICs) or software implementations. They implement both the versions of DES, DES Fast and DES Small and compare their performances. Though DES Small reduces the resource requirements on PLD devices for DES engine, the lack of pipeline in DES Small slows the speed by 10%.

M. Gasser, A. Goldstein, C. Kaufman and B. Lampson [7], in their paper, explains in depth the Digital Distributed System Security Architecture that covers almost all the security specifications for distributed systems. They discuss about the various security goals that are achieved through the use of various security techniques, like confidentiality and integrity achieved using symmetric key cryptography while key distribution, authentication and certification achieved using asymmetric key cryptography. They also talk about creating secure communication channels between sender and receiver and also further say that a symmetric key channel can ensure confidentiality and provide authentication with the use of MAC for integrity. Even though the most popular algorithm for symmetric key encryption is Data Encryption Standard (DES), it is not specified by the architecture. – Symmetric Key used for data exchange as it is faster than asymmetric. Specialized hardware will be required to calculate DES at a speed adequate for data exchange

M. N. Islam, M. M. H. Mia, M. F. I. Chowdhury and M. A. Matin [8], in their paper, talk about providing security to data in motion. They mention that the 64 bits block size and 64 bits key size Data

Encryptions Standard (DES) algorithm is vulnerable to brute force attack, and hence the need for a stronger and larger block size algorithm, the Advanced Encryptions Standard (AES). In this paper, they show the effect in security increment through AES methodology, by proposing an algorithm that is higher secure than Rijndael but comparatively less efficient.

J. Daemen and V. Rijmen [4], in their paper, discuss the Rijndael algorithm's design principles, its advantages and disadvantages. They discuss the need of AES and its widespread implementations in various areas. They also talk about the AES security against practical and academic cryptanalysis methods like statistical attacks, algebraic attacks and the rebound attack and related-key attacks. They say that though DES has been de-standardized, it still approves the use of 3-key triple DES.

W. E. Burr [2], in his paper, mainly discusses about the selection of Rijndael algorithm by National Institute of Standards and Technology (NIST) as the Advanced Encryption Standard from 15 qualifying algorithms. He talks about the limitations of Data Encryption Standard (DES) algorithm and there was a need for it to be replaced. He mentions the fact that DES was designed for hardware implementations and hence poorly suited for software implementations. He mentions the five finalist candidate algorithms – MARS, RC6, Rijndael, Serpent and Twofish; and also the selection process of the final AES. The selecting criteria included general security (the most important criteria), performance and algorithm characteristics, including intellectual property. The main goal was to make the AES to be available worldwide royalty free.

## 4. DES vs. AES – The Final Comparison

Distributed systems can be secured using various mechanisms and in this paper we have reviewed the symmetric key cryptographic approach. Even though symmetric key is used for data exchange, as it is faster than asymmetric key, specialized hardware will be required to calculate DES at a speed adequate for data exchange. Table 1, given below, gives a comparative study of both the algorithms – DES and AES. From the table, it can be observed that AES algorithm provides much more reliability and security than the DES algorithm mainly due to the doubling of the key size of the former algorithm. DES was built mainly for hardware implementations and hence is faster in hardware while AES is faster in software.

**Table1.** DES vs. AES

|  | DES | AES | Ref. |
|---|---|---|---|
| **Block Size (bits)** | 64 | 128 | [4][2][8] |
| **Key Size (bits)** | 56 | 128,192,256 | [4][2][8] |
| **Reliability** | × | ↓ | [6] |
| **Faster** | H/w | S/w | [12][2] |
| **Efficiency** | ↓ | ↑ | [12][2] |
| **Confidentiality** | ↓ | ↑ | [7][8] |
| **Integrity** | ↓ | ↑ | [7][8] |
| **Performance – H/W** | ↑ | ↓ | [4][2] |
| **Performance – S/W** | ↓ | ↑ | [4][2] |
| **Security w.r.t Key Size** | ↓ | ↑ | [4][2] |
| **Royalty Free** | × | ↓ | [2] |

## 5. Conclusion

Out of the various mechanisms available for securing distributed systems, symmetric key cryptography approach has its own benefits over asymmetric key cryptographic technique. In our paper, we have reviewed two of the widely used symmetric key cryptographic technique – AES and DES. From the literatures reviewed of various implementations and analysis of both the algorithms, it can be concluded that AES algorithm has over-shadowed the DES algorithm in many areas. The NIST has also revoked the DES as a standard, even though a variation of it, the 3DES algorithm, is still widely used. The AES algorithm has the main advantage of being royalty free, unlike the DES; so it can be implemented anywhere and extended anyhow. The advantages of the AES algorithm over the DES algorithm can be seen in Table 1 and has also been discussed in the above section. The AES algorithm alone is sufficient to ensure the security wherever it is applied, for a long time to come till various complicated cracking techniques are developed. The AES algorithm has undergone tremendous cryptanalysis for ensuring its robustness and efficiency.

## Authors Profile

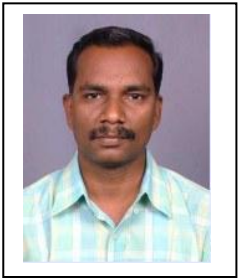

**Prof. Ramesh Babu** is currently working as an Associate Professor at VIT University. He has been associated with VIT University for more than seven years. He did his graduation in Computer Science and Engineering from Pondicherry Engineering College. He has got many publications to his credit and is currently focussing on Distributed Computing and Computer Networks.

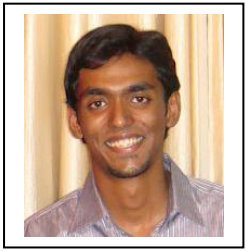

**George Abraham** received his B. E. degree in the year 2011 from Vidyalankar Institute of Technology (VIT) affiliated to Mumbai University (MU), Mumbai, India. He is currently pursuing his M.Tech in Computer Science and Engineering from VIT University, Vellore, India. His areas of interests include Web Services, Distributed Systems and Systems Security.

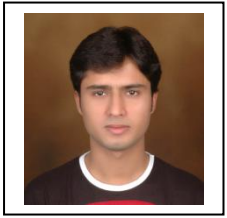

**Kiransinh Borasia** received his B.E degree from Dharmsinh Desai University (DDIT), Nadiad, India. He is currently doing his M.tech in Computer Science and Engineering from VIT University, Vellore, India. His areas of interests include Database management system, Database Tuning, Algorithm.